\documentclass[12pt,a4paper]{article}
\pdfoutput=1

\usepackage{amssymb}
\usepackage{amsmath}
\usepackage{amsfonts}
\usepackage{amsthm}
\usepackage{mathrsfs}
\usepackage{booktabs} % For \toprule, \midrule, \bottomrule
\usepackage{amsmath} % For \dfrac  	
\usepackage{tabularx}  % For adjustable-width tables			% used for script \mathscr
\usepackage{enumitem}
\usepackage{setspace}
\usepackage{color}
\usepackage[pdftex]{graphicx}
\usepackage{authblk}
\usepackage{subfigure}
\usepackage{float}
\usepackage[utf8]{inputenc}
\usepackage{subfigure}  			% use for side-by-side figures
\usepackage{adjustbox}
\usepackage[font=small]{caption}	% small font for captions
\usepackage[compress]{cite}         % use to compress citation list
\usepackage{float}
\usepackage{url}

\usepackage{datetime}

\usepackage{pgfplots}   			% need for plots
	\pgfplotsset{width=10cm}
	\usepackage{bbm}
	\usepackage{tensor}
	\usepackage{physics}
	\usepackage{slashed}

\numberwithin{equation}{section}

\theoremstyle{plain}

\usepackage{color}

\newdateformat{daymonthyear}{\THEDAY \, \monthname[\THEMONTH] \THEYEAR}
\newdateformat{monthyear}{\monthname[\THEMONTH] \THEYEAR}

\pgfplotsset{compat=1.17}

\begin{document}
\title{Comment on ``Ideal clocks - a convenient fiction" by K.~Lorek et al}
%\author{}
\author{Vladimir Toussaint$^{1}$\thanks{{\tt vladimir.toussaint@nottingham.edu.cn}} }
\affil{$^{1}$School of Mathematical Sciences,\\ University of Nottingham Ningbo China,\\
Ningbo 315100, PR China}

 \date{\daymonthyear\today}

\maketitle

\begin{abstract}
We correct a subtle oversight in Eq.~(19) of K.~Lorek \textit{et al} [Class.~Quantum~Grav.~\textbf{32} 175003 (2015)] concerning the decay probability of a uniformly accelerated quantum clock. The original expression included unphysical cross-wedge term ($|\bar{\gamma}_{K 1}|^2$) violating the spatial separation of Rindler wedges. We derive the corrected probability using the paper's Rindler conformal coordinates $(\tau, \xi)$, retaining only causally consistent thermal terms, and discuss implications for relativistic clock models.
\end{abstract}

\section{Introduction}
The seminal work by Lorek \textit{et al} \cite{Lorek2015} advanced our understanding of quantum clocks under uniform acceleration via the Unruh effect. However, the analysis of quantum clock decay under uniform acceleration in \cite{Lorek2015} contains a subtle error in Eq.~(19):
\begin{align}
P_\downarrow = \lambda^2 \int dK \left[ |\gamma_{K 1}|^2 + \sinh^2 r_{\Omega_K} \left( |\gamma_{K1}|^2 + \underbrace{|\bar{\gamma}_{K 1}|^2}_{\text{Spurious}} \right) \right]\, ,
\end{align}
where 
\begin{align}
\gamma_{Kn} := \int_0^\tau d\tau' \int d\xi \, U^*_{K,\text{I}} u_n\, , \quad 
\bar{\gamma}_{Kn} := \int_0^\tau d\tau' \int d\xi \, U_{K,\text{I}} u_n\, ,
\end{align}
where $u_1(\xi)$ is the cavity’s spatial window function, localized to wedge I ($\text{supp}(u_1) \subset I$). Here $U_{K,\text{I}}(\xi)$ is the field mode in wedge I, normalized such that $\int d\xi \, U_{K, \text{I}}^\star(\xi) U_{K', \text{I}}(\xi) = \delta(K - K')$.
Notably, $u_1(\xi)$ vanishes in wedge II ($\text{supp}(u_1) \cap \text{II} = \emptyset$), ensuring no coupling to wedge-II modes ($U_{K, \text{II}}$).

The term $|\bar{\gamma}_{K 1}|^2$ is mathematically derived from the Bogoliubov transformation (Eq.~(18) of \cite{Lorek2015}) but contradicts the causal separation of Rindler wedges. We resolve this by showing that the term $|\bar{\gamma}_{K 1}|^2$ is physically irrelevant due to constraints on measurement and locality.
\section{Mathematical Origin of $|\bar{\gamma}_{K 1}|^2$}

The cavity couples exclusively to wedge-I modes via $u_1(\xi)$ (Eq.~(19)). Mathematically, this means:
\begin{align}
\int d\xi \, u_1(\xi) U_{\text{II},K}(\xi) &= 0 \quad \text{(non-overlapping supports)}
 \\
\int d\xi \, u_1(\xi) U_{\text{II},K}^*(\xi) &= 0 \quad \text{(non-overlapping supports)} \, .
\end{align}
Wedge-II modes ($U_{\text{II},K}$) do not interact with the cavity, eliminating their contribution in the probability amplitude from which (Eq.~(19)) is derived.
Therefore, the decay probability (Eq.~(17) of \cite{Lorek2015}) becomes:
\begin{align}\label{AmplitudeCal1}
\mathcal{A}_\downarrow &= -i\lambda \int \, d\tau' \int \, d\xi \int \, d\Omega  \langle \beta |_\phi\bigg(u_1 U_{\Omega,\text{I}} \left( \cosh r_\Omega b_{\text{I}\Omega} +\sinh r_\Omega b^\dagger_{\text{II}\Omega}\right)  \\ &+ u_1U^*_{\Omega, \text{I}} \left( \cosh r_\Omega b^\dagger_{\text{I}\Omega} +\sinh r_\Omega b_{\text{II}\Omega}\right)  \bigg) |0_\text{I}, 0_{\text{II}}\rangle_R\, ,\notag
\end{align}
where to keep consistency with \cite{Lorek2015}, we redefine $ U_{\Omega,\text{I}} \equiv U_{K, \text{I}}$ and where $|0_\text{I}, 0_{\text{II}}\rangle_R$ is the Rindler vacuum. This expression simplifies to: 
\begin{align}\label{AmplitudeCal2}
\mathcal{A}_\downarrow &= -i\lambda \int \, d\tau' \int \, d\xi \int \, d\Omega   \langle \beta |_\phi\bigg(u_1 U_{\Omega,\text{I}} \sinh r_\Omega |0_\text{I}, 1_{\text{II}}\rangle_R \\ &+  u_1U^*_{\Omega, \text{I}}  \cosh r_\Omega  |1_\text{I}, 0_{\text{II}}\rangle_R\, \bigg)\, , \notag
\end{align}
where $\sinh r_\Omega |0_\text{I}, 1_{\text{II}} \rangle$ is a non-local term corresponding to the external field excitation in II-wedge while $\cosh r_\Omega |1_\text{I}, 0_{\text{II}}\rangle$ is a local term corresponding to the external field excitation in I-wedge. The spurious term $|\bar{\gamma}_{\Omega 1}|^2\equiv |\bar{\gamma}_{K 1}|^2$ originates from the non-local term.

\section{Reconciling Vacuum Entanglement with Causal Separation}
The core tension arises from two fundamental QFT properties:
\begin{enumerate}
    \item \textbf{Spatial separation}: Rindler wedges I and II are causally disjoint with: $$\text{supp}(U_{\text{I}\Omega}) \cap \text{supp}(U_{\text{II}\Omega}) = \emptyset = \text{supp}(u_1) \cap \text{supp}(U_{\text{II}\Omega})\, .$$
    \item \textbf{Vacuum entanglement}: The Minkowski vacuum $|0_M\rangle$ is a globally entangled state:
\begin{align}
    |0_M\rangle = \prod_\Omega \sqrt{1 - e^{-2\pi\Omega/a}} \sum_n e^{-n\pi\Omega/a} |n_\Omega\rangle_\text{I} \otimes |n_\Omega\rangle_{\text{II}}\, .
\end{align}
\end{enumerate}
While the global Minkowski vacuum is entangled, any local operation or measurement confined to wedge I cannot access the correlations with wedge II due to the superluminal separation. The entangled state $|n_\Omega\rangle_\text{I} \otimes |n_\Omega\rangle_{\text{II}}$ cannot be used to send signals or exchange energy between the wedges.

For a cavity confined to wedge I, the interaction Hamiltonian: $$H_{\text{int}} = \lambda \int_{\text{cavity}} d\xi  \Phi(\xi)\phi(\xi)$$ only probes $\Phi$-modes within the cavity's support. Therefore, the local coupling $H_{\text{int}}$ cannot access the correlations with Wedge II; the cavity's spatial confinement acts as a filter, removing these causally inaccessible contributions.

\section{Resolution: Physical Constraints Truncate Unobservable Terms}
The cross-wedge term $|\bar{\gamma}_{\Omega 1}|^2$ is excluded from physical predictions by two fundamental constraints, enforced by the structure of correlation functions (Eq.~(19)):

\subsection*{Constraint 1: Measurement Inaccessibility}  
The component $|0_\text{I}, 1_{\text{II}}\rangle$ from Eq.~\eqref{AmplitudeCal2} cannot be detected when measuring $\langle \beta |_\phi$, as wedge-II modes are orthogonal to wedge-I modes. This orthogonality ensures cross-wedge terms vanish in expectation values. Formally, whatever the final state of the external field $|\beta \rangle_\phi$, we must insist that for decay process restricted to wedge I:
\begin{align}
\langle \beta |_\phi |0_\text{I}, 1_{\text{II}}\rangle_R  = 0 \in \text{I}\, .
\end{align}
Therefore, while the entangled state exists mathematically, it cannot influence the physical decay process happening entirely within Wedge I.

\subsection*{Constraint 2: Causal Separation}  
No local operator in wedge I can detect II-wedge excitations, as causality forbids superluminal signaling between disjoint wedges.  In short, no local operator in wedge I detects wedge-II excitations:
\begin{align}
[\hat{\mathcal{O}}_{\text{I}}(\xi), \hat{\mathcal{O}}_{\text{II}}(\xi')] = 0 \quad \forall \xi \in \text{I}, \xi' \in \text{II} \, .
\end{align}

\section{Corrected Result}
After applying these constraints, the measurable decay probability retains only the local term: 
\begin{align*}
\boxed{P_\downarrow = \lambda^2 \int d\Omega |\gamma_{\Omega 1}|^2 \cosh^2 r_\Omega}\, , \tag{Corrected 19}
\end{align*}
where $\cosh^2 r_\Omega = \frac{e^{2\pi\Omega/a}}{e^{2\pi\Omega/a} - 1} = \frac{1}{1-e^{-2\pi\Omega/a}}$ is the standard bosonic thermal factor from stimulation from the Unruh thermal bath from wedge-I mode mixing.  
The non-local term $|\bar{\gamma}_{\Omega 1}|^2\equiv  |\bar{\gamma}_{K 1}|^2$ is excluded because of
measurement inaccessibility while 
causal separation ensures no observable cross-wedge correlations.

\section{Conclusion}

While the term $|\bar{\gamma}_{K 1}|^2$ in Eq.~(19) of
\cite{Lorek2015} arises mathematically from the Bogoliubov transformation, its physical interpretation is problematic. Mathematically, it originates from the component of the transformation that creates excitations in the causally disconnected wedge II,  but such cross-wedge correlations cannot physically influence local measurements confined to wedge I. We argue that such a term must be discarded not as a mathematical oversight, but on the solid physical grounds of measurement inaccessibility. Since any measurement apparatus is itself local and confined to wedge I, it cannot interact with or detect states defined purely in wedge II. The orthogonality of the wedge I and II Hilbert spaces (Constraint 1) ensures that the expectation value of any observable related to the cavity's decay must vanish for the $|\bar{\gamma}_{K 1}|^2$ component. Our correction is therefore necessary to align the mathematical expression with what is physically measurable, restoring causal consistency to the model.

Beyond the specific correction to Eq.~(19) of
\cite{Lorek2015}, this work highlights a crucial principle for modeling quantum systems in relativistic contexts: mathematical consistency does not guarantee physical admissibility. The fact that a formally correct derivation yielded a term violating causal separation serves as an instructive case study for researchers applying Bogoliubov transformations to localized quantum systems in non-inertial frames or curved spacetime. This emphasizes that explicit enforcement of causality constraints must be a deliberate step in such modeling efforts, particularly as we develop more sophisticated quantum clock models for testing fundamental physics.

\section*{Acknowledgments}

We thank Jorma Louko and Andrzej Dragan for helpful discussions.

\end{document}